%% file: smuggling.tex
\documentclass[12pt]{article}
\usepackage{amsmath, amssymb, amsthm}
\usepackage{graphicx, url}
\usepackage{rotating}
\usepackage{multicol}
\usepackage{enumitem}
\usepackage{subfig}
\usepackage{booktabs}
\usepackage{epigraph}
\usepackage[all]{xy}
\usepackage{verbatim}
\usepackage{natbib}
\usepackage{tabularx}
\usepackage[page]{appendix}

\newcommand{\blind}{0}
\newcommand{\submission}{0}

\usepackage[T1]{fontenc}
\if0\submission{}
\usepackage{cochineal}
\usepackage[vvarbb]{newtxmath}
\fi
\usepackage{pgf, tikz}
\usetikzlibrary{positioning}
\usetikzlibrary{arrows,automata}
\newcommand{\E}{\mathbb{E}}

\renewcommand{\P}{\mathbb{P}}

\newcommand{\bX}{\mathbf{X}}

\newtheorem{assumption}{Assumption}

\newcommand{\indep}{\perp\!\!\!\perp}

  \newtheorem{proposition}{Proposition}

\if0\blind{}
  \def\myauthor{Matthew Blackwell, Ruofan Ma, and Aleksei Opacic}
\else
\def\myauthor{}
\fi
\def\mytitle{Assumption Smuggling in Intermediate Outcome Tests of Causal Mechanisms}
\def\mykeywords{causal inference; mechanisms}

\if1\submission{}
\usepackage{xr}
\usepackage{setspace}
\usepackage{caption}
\else
  \usepackage{fullpage}
\fi

\definecolor{gray}{rgb}{0.459,0.438,0.471}
\definecolor{crimson}{rgb}{0.6,0,0}

\usepackage[unicode,
plainpages=false, 
pdfpagelabels, 
bookmarksnumbered,
pdftitle={\mytitle}, 
pdfauthor={\myauthor},
pdfkeywords={\mykeywords},
colorlinks=true,
citecolor=crimson, 
linkcolor=crimson, 
urlcolor=crimson]{hyperref} 

\usepackage{cleveref}
\title{{\bf \mytitle}\if0\blind{}\thanks{Working paper, comments welcome. Thanks to David Broockman, Thad Dunning, Felix Elwert, Chad Hazlett, Nahomi Ichino, Gary King, Adeline Lo, Cyrus Samii, Maya Sen, Anton Strezhnev, Jessica Weeks, and participants at the University of Wisconsin Models, Experiments, and Data Workshop for helpful comments and conversations.}\fi
}
\if0\blind{}
\author{Matthew Blackwell\thanks{Department of Government and Institute for Quantitative Social Science, Harvard University.\ web: \mbox{\url{http://www.mattblackwell.org}}, email: \texttt{\href{mailto:mblackwell@gov.harvard.edu}{mblackwell@gov.harvard.edu}}}\and
  Ruofan Ma\thanks{Department of Government, Harvard University.\ email: \texttt{\href{mailto:ruofan_ma@g.harvard.edu}{ruofan\_ma@g.harvard.edu}}}\and
  Aleksei Opacic\thanks{Department of Sociology, Harvard University.\ web: \mbox{\url{https://alekseiopacic.github.io}}, email: \texttt{\href{mailto:aopacic@g.harvard.edu}{aopacic@g.harvard.edu}}}%
}
\fi

\date{
  \today \if1\submission{}\\ Word Count: 11,838\fi
}

\begin{document}

\maketitle

\begin{abstract}
\noindent Political scientists are increasingly interested in assessing causal mechanisms, or determining not just \emph{if} a causal effect exists but also \emph{why} it occurs.  Even so, many researchers avoid formal causal mediation analyses due to their stringent assumptions, instead opting to explore causal mechanisms through what we call \emph{intermediate outcome tests}. These tests estimate the effect of the treatment on one or more mediators and view such effects as suggestive evidence of a causal mechanism. In this paper, we use nonparametric bounding analysis to show that, without further assumptions, these tests can neither establish nor rule out the existence of a causal mechanism. To use intermediate outcome tests as a falsification test of causal mechanisms, researchers must make a very strong but rarely discussed monotonicity assumption. We develop a way to assess the plausibility of this monotonicity assumption and estimate our bounds for two recent experiments that use these tests.

\end{abstract}
%TC:endignore
\clearpage

\if1\submission{}
\setstretch {1.618}
\else
\setlength{\baselineskip}{1.57\baselineskip}
\fi

\section{Introduction}

Over the last few decades, causal inference has become a bedrock of quantitative research in political science, with greater care devoted to defining causal quantities of interest and stating and assessing the required assumptions. For good reasons, the main focus of this causal turn has been on establishing the existence (or nonexistence) of a causal effect on some interesting outcome. But scholars are often even more interested in \emph{why} a causal effect exists as much as \emph{if} a causal effect exists, leading to the proliferation of causal mechanism tests in empirical political science. 

\begin{figure}[t]
\vspace{1cm}
\begin{center}
\includegraphics[width=.6\textwidth]{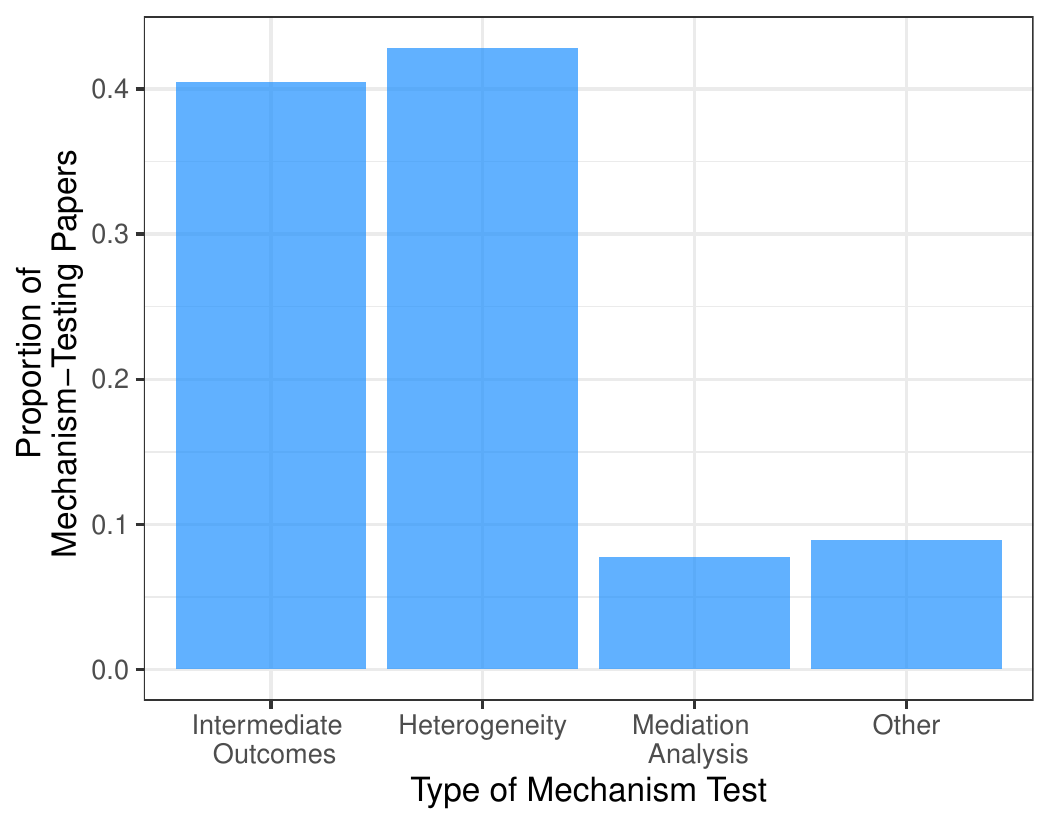}
\caption{Proportion of papers analyzing causal mechanisms using different approaches from top journals in political science (APSR, AJPS, JOP), 2022–2023.}
\label{fig:motivation}
\end{center}
\end{figure}

There has long been disagreement about the proper way to test for causal mechanisms, leading to a fractured landscape and difficulty understanding how different tests relate. From a formal point of view, causal mechanisms have been most closely linked to mediation analysis, wherein the overall average treatment effect is decomposed into its direct (net of a mediating variable of interest) and indirect effects \citep{Pearl01, ImaKeeYam10, ImaKeeTin11}. 
Some scholars, however, object to causal mediation analysis because of the strong assumptions it requires \citep{GreHaBul10, acharya2016explaining, BulGre21, CalDunTun24}. In particular, identification of the average direct and indirect effects relies on unverifiable assumptions that (i) the mediator is as-if randomized condition on pretreatment covariates, and (ii) there are no other post-treatment variables that confound the mediator-outcome relationship. This skepticism is reflected in the relative lack of causal mediation analyses in recent empirical political science articles. We analyzed 487 empirical papers published in the \emph{American Political Science Review}, the \emph{American Journal of Political Science}, and \emph{The Journal of Politics} between 2022 and 2023 and found that 161 (33\%) provided at least one quantitative test for causal mechanisms. As shown by Figure~\ref{fig:motivation}, however, of these 161 papers, only 16 (10\%) employed a formal causal mediation analysis. 

If these studies do not employ formal techniques of causal mediation, how do they assess causal mechanisms? Figure~\ref{fig:motivation} shows that researchers commonly use an alternative approach to evaluating causal mechanisms, which we call in this paper we call \emph{intermediate outcome tests} or IOTs. In these tests, researchers estimate the causal effect of the treatment on one or more potential mediators, often using the same research design used to estimate the main effect of the treatment on the outcome \citep{GreHaBul10, CalDunTun24}. Under this approach, if researchers detect an average treatment effect on a mediator, this mediator is considered part of a potential ``mechanism'' for the treatment effect on outcome. IOTs have clear, intuitive appeal since it seems to follow that, for $M$ to mediate the relationship between treatment $A$ and outcome $Y$, there must be an effect of $A$ on $M$ and then of $M$ on $Y$. The (often implicit) argument for the IOT approach is that an average effect of $A$  on $M$ is a \emph{necessary} and sometimes sufficient condition for $M$ to be a causal mechanism.  Thus, it is no surprise that this simple ad hoc approach is one of the dominant ways to estimate causal mechanisms, accounting for roughly 40\% of the papers in our literature review.

In this paper, we show that, in an experimental context, intermediate outcome tests under randomization of treatment alone can neither establish nor rule out the existence of an indirect effect of treatment on the outcome through that mediator. Papers that imply otherwise engage (often unknowingly) in what we call \emph{assumption smuggling}---presenting suspect conclusions that rely on strong hidden assumptions alongside well-grounded findings that depend on the more reasonable, explicitly stated assumptions.  We establish the IOT's inability to inform mechanisms using sharp nonparametric bounds that contain all values of the indirect effect consistent with the observed data and the maintained assumptions. Under randomization of treatment alone, these bounds are never informative about the sign of the indirect effect for two reasons. First, the average effect of the treatment on the mediator might mask heterogeneous treatment effects that are important to the causal mechanism. Second, establishing indirect effects requires strong evidence or assumptions about the effect of the mediator on the outcome at the individual level. 

To help researchers navigate this, we further introduce an assumption that, though very rarely discussed in practice, does enable the IOT procedure to have a meaningful interpretation. Specifically, we investigate a \emph{monotonic mediator response} assumption wherein the units’ value of the mediator can only be affected by treatment in one direction. Under this assumption, we show that a lack of treatment effect on the mediator can, in fact, rule out indirect effects through that mediator. We also derive the nonparametric sharp bounds for the average indirect effect under this assumption, showing that IOTs can ``approximately'' rule out mechanisms in that small (zero) effects of the treatment on the outcome imply small (zero) indirect effects. Of course, monotonicity is a strong, untestable assumption that will frequently fail to hold in applied practice, and there is no design-based way to ensure it holds. Thus, it is critical for researchers to state this assumption clearly and to interrogate it if they wish to use IOTs to assess causal mechanisms at all. We present one falsification test for this assumption that relies on nonparametric tests of effect heterogeneity, such as causal forests. 

Can IOTs ever establish the sign of an indirect effect? In some substantive settings, researchers may care more about the directional relations between the treatment and the mediator, rather than pinpointing the exact magnitude of the effect. We also show that, unfortunately, it is nearly impossible to establish the sign of an indirect effect even under very strong assumptions. In particular, we show that even if we \emph{assume} a positive average effect of the mediator on the outcome, the sharp nonparametric bounds on the indirect effect will never be informative about the sign of that effect. Thus, while the monotonic mediator response assumption may allow researchers to rule out potential pathways, no intermediate outcome test can establish the simple \emph{existence} of a causal mechanism even if we make heroic assumptions about the mediator-outcome relationship. 

The contributions of this paper are twofold. First, we highlight for applied researchers that this \textit{ad hoc}, widely used approach to analyze causal mechanisms actually relies on strong and usually unstated assumptions; relaxing the assumptions, as we show, can render substantive conclusions invalid. Second, by clarifying the assumptions needed, we provide guidance for interpreting these IOTs in terms of the substantive application.  The takeaway message of this paper reinforces what has been argued elsewhere: establishing causal mechanisms is difficult and requires strong assumptions regardless of the approach. Researchers may not avoid this difficulty even when using IOTs.

Our work contributes to a broader literature on causal mechanisms and (in)direct effects in multiple fields. Many scholars have formalized the concept of direct and indirect effects in terms of potential outcomes and have derived assumptions that can point identify these quantities \citep{Pearl01, ImaKeeYam10}. Most similar to our own work is \cite{imai2013experimental}, which derived nonparametric bounds for indirect effects under several different experimental designs, and \cite{Glynn12}, which showed how individual-level heterogeneity in treatment effects renders many indirect effect tests ineffective. We expand on their results by showing how different causal assumptions can change the bounds for a single experiment. Several studies have proposed alternative ways to assess causal mechanisms, such as implicit mediation \citep{BulGre21} and effect heterogeneity \citep{fu2024heterogeneous}. However, these papers show that these alternatives also rely on strong assumptions. An alternative definition of causal mechanisms centers on a quantity known as the ``controlled direct effect'' (CDE), which captures the average direct effect of a treatment variable while holding the mediator fixed at a given value \citep{goetgeluk2008estimation,acharya2016explaining, acharya2018analyzing, zhou2019regression, blackwell2022telescope}. While identifying the CDE requires weaker assumptions than mediation, many mediation-skeptical scholars object to these approaches because they require us to identify the effect of the mediator on the outcome.\footnote{Some recent work investigates how to handle multiple mediating variables \citep{imai2013identification, vanderweele2014effect, daniel2015causal, zhou2023tracing}. These approaches typically rely on stronger identification assumptions than the single-mediator case and are thus less common in political science. For these reasons, we focus on the single-mediator setting.}

Two recent papers are directly related to our setting. In an applied setting of human rights treaties, \cite{StrKelSim21} explores a sensitivity analysis procedure for estimating indirect effects under a set of linear models, which is a familiar setting for many empirical scholars. Our results, on the other hand, are fully nonparametric and do not depend on functional form assumptions. \cite{KwoRot24} proposes a way to test whether a given mediator explains all of the effect of treatment, which they call the ``sharp null hypothesis of full mediation,'' and also highlight the importance of a monotonicity assumption on the effect of treatment on the mediator. Their approach focuses on a novel quantity of interest (the sharp null), whereas we show what researchers can learn about more traditional causal mechanism estimands. 

This paper proceeds as follows. In Section~\ref{sec:motivating}, we discuss the setting of our two empirical applications, showing how researchers use IOTs in practice. Since the phrase ``causal mechanisms’’ means different things to different scholars, we next lay out our preferred notation and definitions for causal effects and causal mechanisms in Section~\ref{sec:notation}. In Section~\ref{sec:iot_analysis}, we formally analyze intermediate outcomes tests using a principal stratification approach and derive sharp nonparametric bounds under different assumptions. Section~\ref{sec:applications} applies these bounds to two recent studies, showing the limited information these tests provide about indirect effects. Section~\ref{sec:why} discusses how scholars might consider mechanisms when they cannot meet the assumptions we discuss in this paper. Section~\ref{sec:conclusion} concludes and describes avenues for future research. 

\section{Two Motivating Applications}\label{sec:motivating}

\subsection{Reducing Outgroup Prejudice}\label{sec:motivating-kalla}
An important literature in political psychology examines intergroup prejudices and evaluates how different individual-oriented interventions might durably reduce such prejudices. An approach that has garnered empirical support in a growing literature is the use of interpersonal conversations 
\citep{galinsky2000perspective, bruneau2012power, broockman2016durably, adida2018perspective, simonovits2018seeing, audette2020personal, lowe2021types, williamson2021family}, but an unresolved question in this recent work is which ``narrative strategies'' are most effective at reducing prejudice. For example, an interpersonal conversation that emphasizes taking the perspective of an outgroup member (``perspective taking'') may yield different outcomes from one that simply describes the negative experiences of the outgroup (`` perspective getting'').

To evaluate the effects of distinct forms of interpersonal conversation on reducing prejudice, \cite{kalla2023narrative} implement three field experiments alongside a survey experiment. In the field experiments, volunteer canvassers went door to door (or phoned houses, in the case of the third) to a sample of registered voters across multiple states, engaging in conversation about unauthorized immigrants. Each respondent was randomly assigned to one of several narrative strategies. In the survey experiment, researchers showed respondents a picture of an outgroup member and asked them to engage in an exercise corresponding to different narrative strategies. 

Across these experiments, the authors find narratives that simply describe the outgroup's experience of discrimination consistently and durably reduce prejudicial attitudes as much as the narratives that additionally ask respondents to imagine how they may feel if they had a similar experience. But why does perspective-getting reduce prejudice? Addressing this question is important not only because of its theoretical dividends but also because it offers important insights for practitioners. Given that prejudice interventions are often expensive and unscalable, discerning \textit{how} perspective-getting reduces exclusionary attitudes facilitates the development of more cost-effective interventions that target these mechanisms. 
The authors implement an IOT to investigate mechanisms, showing positive, statistically significant effects of perspective-getting narratives on a handful of potential mediators, and argue that intervention ``activates multiple mechanisms that have been found to reduce prejudice'' \citep[p. 188]{kalla2023narrative}.

\subsection{Transitional Justice Museums and Support for Democracy}\label{sec:motivating-museums}

Many scholars have credited transitional justice policies with facilitating democratic consolidation, reconciliation, and peace-building in post-conflict societies \citep{de2001politics, horne2014impact, nalepa2010skeletons}.  Less is known, however, about how such policies operate at the individual level and through what mechanisms. In a field experiment in Chile, \cite{balcells2022transitional} study one particular instance of transitional justice policy: museums. Researchers randomized students into either a control condition or a treatment condition consisting of an hour-long tour of a museum memorializing victims of General Augusto Pinochet’s dictatorship. The authors then fielded a survey about beliefs about democracy and found that treated students ``display greater support for democratic institutions, are more likely to reject institutions associated with the repressive period, and are more supportive of restorative transitional justice policies" \citep[][p. 496]{balcells2022transitional}. 

Identifying how such an intervention increases support for democratic institutions could help inform the design of other, potentially less costly, policies. The authors implement an IOT procedure using a battery of emotion-oriented questions to understand the mechanisms underlying their estimated effects. They find that treated students demonstrate higher levels of emotion in both a positive (e.g., feeling stimulated, inspired, interested) and negative (e.g., feeling tense, nervous, embarrassed) sense. The authors take their IOT results as evidence that the ``emotional experience" elicited by a museum visit is an important pathway that explains its promotion of democratic sentiment. 

Lastly, while we do not evaluate all mediators and outcomes proposed in the original studies given their richness, we highlight a selected set of variables in line with the main theoretical arguments of the original authors. Table~\ref{tab:motivating_examples} lists the mediators and outcomes in each of the examples discussed in this section that we will be focusing on in the application section. Further discussions on why we select these variables can be found in Section~\ref{sec:applications}.

\begin{table}[ht]
\centering
\begin{tabularx}{\textwidth}{|X|X|X|X|}
\hline
\textbf{Example} & \textbf{Treatment} & \textbf{Mediator} & \textbf{Outcome} \\ \hline
\cite{kalla2023narrative} & Perspective-getting (i.e., reading a story about an outgroup member and then summarizing the story) during canvassing & Attributional thinking; Entativity; Reactive empathy; Self-outgroup merging & Reduced prejudice against transgender people and undocumented immigrants \\ \hline
\cite{balcells2022transitional} & Museum tour about victims of Pinochet’s dictatorship & Negative emotions elicited during the tour (e.g., feeling tense, scared, guilty, etc.)& Increased support for democratic institutions in Chile\\ \hline
\end{tabularx}
\caption{Summary of Treatments, Mediators, and Outcomes in the Motivating Applications}
\label{tab:motivating_examples}
\end{table}

\section{Causal Mechanisms: Notation and Quantities of Interest}\label{sec:notation}

These two studies illustrate how many political scientists generate substantive conclusions about ``causal mechanisms,'' the meaning of which is the source of some debate in the methodological and statistical literature. We explore the interpretation of mediation-style indirect effects as one measure of a causal mechanism.  To explore this, we focus on a simple yet common empirical setting in the social sciences: the effect of a binary treatment on a binary outcome with a binary mediator. Obviously, this does not capture all settings in political science, but many of the core ideas we present generalize to other settings easily. In this section, we introduce basic notation and some well-understood issues in the identification of direct and indirect effects; this sets the stage for the paper’s contributions in the following section, in which we formalize the assumptions necessary for the IOT approach to be informative about the mechanism.

Let $A_i \in \{0, 1\}$ be a binary treatment variable, $M_i \in \{0, 1\}$ be the binary mediator, and $Y_i \in \{0, 1\}$ be the binary outcome. In some cases, there may also be a vector of pretreatment covariates, $\bX_i$, to aid identification or make estimates more efficient. We refer to the $(A_i, M_i, Y_i, \bX_i)$ as the observed data and assume the vector is an independent and identically distributed draw from some population distribution. 

We use the potential outcomes framework for causality. Let $Y_i(a, m)$ be the outcome that unit $i$ would have if (potentially counter to fact) they received treatment $A_i = a$ and mediator $M_i = m$. We make the usual consistency assumption that ties these potential outcomes to observed outcomes:
$$
Y_i = Y_i(a,m) \qquad \text{if} \quad A_i = a, M_i = m.
$$
In mediation analyses, we are also concerned about the effect of treatment on the mediator, so we define potential outcomes for the mediator, $M_i(a)$ as well. In this framework, causal effects are contrasts between different potential outcomes. 

With this notation, it is possible to define potential outcomes that involve ``natural'' values of the mediator, such as $Y_i(1, M_i(0))$, which is the outcome that unit $i$ would take if it was assigned to treatment but had received its mediator value under the control condition. These types of potential outcomes are important to mediation, but they create challenges in estimation and interpretation due to their ``cross-world'' nature. In particular, a unit cannot simultaneously be subject to treatment and control conditions at the same time, so $Y_i(1, M_i(0))$ is not observable even in principle.

The first causal question in this setting is generally ``what is the overall effect of treatment?'' To this end, the potential outcome is just setting the treatment as $Y_i(a) = Y_i(a, M_i(a))$, and we define the average treatment effect (ATE) as 
$$
\tau = \E[Y_i(1) - Y_i(0)].
$$
Similarly, we define the average treatment effect on the mediator (ATM) as 
$$
\alpha = \E[M_i(1) - M_i(0)].
$$
Finally, the average natural indirect effect (ANIE) for a given treatment level is defined as
\begin{equation}
   \delta(a) = \E[Y_i(a, M_i(1)) - Y_i(a, M_i(0))].
\end{equation}
The ANIE captures the ``path-specific'' idea of causal mechanisms since it is the effect of the mediator induced by a change in treatment status, holding the direct effect of treatment constant. In a similar spirit, we define the average natural direct effect (ANDE) as 
\begin{equation}
    \zeta(a) = \E[Y_i(1, M_i(a)) - Y_i(0, M_i(a))],
\end{equation}
which is the direct effect of the treatment holding the mediator to its natural value under treatment level $a$. An attractive property of these mediation quantities is that they decompose the ATE,
\begin{equation}
    \tau = \delta(a) + \zeta(1 - a),
\end{equation}
allowing a researcher to determine how much of an overall effect is due to the mediator or other factors \citep{Pearl01, ImaKeeYam10}. 

\subsection{Identifying assumptions}

The identification of the above causal effects from observational data requires assumptions. In this paper, we assume a setting where researchers believe the main effects of treatment are identified but are more skeptical of assumptions about the relationship between the mediator and the outcome. We begin with the core identifying assumption of an experimental study. 

\begin{assumption}[Randomization]
\label{assm:randomization}
For all $a,a', m$, $\{Y_i(a, m), M_i(a')\} \indep A_i \mid \bX_i$.
\end{assumption}
%% TODO: add discussion 
Random treatment assignment in the experimental setting can justify Assumption~\ref{assm:randomization}. This assumption is more difficult to justify in observational studies, but a large body of work focuses on so-called ``natural'' experiments that could make this assumption believable. Under this assumption, we can identify the ATE as the average of covariate-specific effects, 
$$
\tau = \E\left[\E[Y_i \mid A_i = 1, \bX_i] - \E[Y_i \mid A_i = 0, \bX_i]\right].
$$

This randomization assumption encodes our belief in the basic design of an empirical paper. However, identifying the direct and indirect effects relies on stronger assumptions about the relationship between the mediator and the outcome. In particular, \cite{Pearl01} and \cite{ImaKeeYam10} have shown that we can identify the ANIE and ANDE by assuming that the mediator is also as-if randomized conditional on treatment and the covariates. 

\begin{assumption}[Mediator as-if randomization]
\label{assm:m_randomization}
    $Y_i(a', m) \indep M_i(a) \mid A_i = a, \bX_i$ 
\end{assumption}

Many authors have expressed skepticism about Assumption~\ref{assm:m_randomization} since it requires that there are no (measured \emph{or} unmeasured) post-treatment confounders for the relationship between the mediator and the outcome.\footnote{One alternative approach that has garnered some attention in recent years involves estimating a quantity known as the controlled direct effect (CDE), which captures the causal effect of a treatment when the mediator is fixed to a given value. This quantity is attractive because it is identified under a weaker assumption than Assumption \ref{assm:m_randomization}, one that allows for the existence of observed ``intermediate confounders'' (confounders that are affected by the treatment and which affect the mediator and outcome). A drawback of this approach is that it does not quantify the extent of mediation through $M_i$ and only enables a researcher to rule out the existence of alternative mediators other than $M_i$  (see \cite{acharya2016explaining}).} This condition is often difficult to sustain when the mediator is observed at its natural value after treatment, as is commonly the case. Indeed, while many are willing to believe the identifying assumptions for treatment, assuming the same condition for the mediator is much less plausible.

\subsection{Assumption smuggling}

The rest of this paper will assume the position of the mediation skeptic who is willing to believe Assumption~\ref{assm:randomization} but not Assumption~\ref{assm:m_randomization}. We will show what these mediation-skeptical researchers can learn about causal mechanisms and indirect effects from intermediate outcome tests alone. 

We define \emph{assumption smuggling} to be the (perhaps unintentional) practice of presenting evidence for a claim that only tests the claim under additional assumptions that are strong and undisclosed. (We limit this definition to strong assumptions to differentiate this practice from omitting regularity conditions like the existence of fourth moments or other technicalities.) The disclosure requirement for an assumption may be satisfied by common knowledge of an assumption (a linear functional form, perhaps) or by explicit disclosure when presenting evidence for the claim.

Empirical studies often contain a section on ``mechanisms'' with empirical tests that purport to elucidate why a causal effect exists without laying out the assumptions these tests require. These mechanism sections are often a prime example of assumption smuggling: attempting to maintain nominally weak assumptions with empirical tests that actually require a stronger set of unmentioned assumptions. Of course, like with unattended luggage in an airport, most instances of assumption smuggling are unintentional.  Parametric models commonly used in political science often embed many strong assumptions, and many of us have built our statistical intuitions from that foundation. But one benefit of the recent ``credibility revolution'' has been to encourage researchers to be more explicit about the assumptions of their research designs. Our goal is to help mediation-skeptic scholars who wish to test causal mechanisms do so with a clear articulation of what exactly they are assuming. 

\section{Intermediate Outcome Tests}\label{sec:iot_analysis}

\subsection{What are IOTs and how are they interpreted?}

We now introduce the first common test for causal mechanisms in the empirical literature: the intermediate outcome test. The logic of an IOT is quite compelling.  We assume that a researcher believes Assumption~\ref{assm:randomization} to justify interpreting the estimated effect of $A_i$ on $Y_i$ as causal. That same assumption will also justify estimating the causal effect of treatment on the mediator with the same basic strategy:
$$
\alpha = \E[M_i(1) - M_i(0)] = \E\left[\E[M_i \mid A_i = 1, \bX_i] - \E[M_i \mid A_i = 0, \bX_i]\right].
$$
IOTs then proceed to interpret estimates of this ATM in terms of how it relates to causal mechanisms, generally, or average indirect effect specifically. There are two interpretations of the IOT results that differ in their strength:
\begin{enumerate}
    \item \textbf{Falsification interpretation}: no ATM ($\alpha = 0$) implies $M_i$ is likely not part of an indirect effect/causal mechanism. 
    \item \textbf{Establishment interpretation}: a non-zero ATM ($\alpha \neq 0$) is suggestive evidence that $M_i$ is part an indirect effect/causal mechanism.
\end{enumerate}
To be clear, the logic of IOTs is often left implicit in papers that deploy them, but their interpretation usually falls into one (or both) of these two buckets. 

The falsification interpretation views IOTs as ``necessary but not sufficient evidence'' for causal mechanisms. In our empirical application on prejudice reduction, the authors find that the perspective-getting intervention has a negative overall treatment effect on prejudicial attitudes, or $\widehat{\tau}$ < 0. We might suspect that a measure of empathy for the outgroup could be part of the causal mechanism here in the sense that perspective-getting increases empathy, which in turn decreases prejudice. Taken together, these effects would imply a negative indirect effect that explains the ATE. The falsification interpretation says that if we do not see an average effect of the perspective getting on empathy, then empathy is unlikely to explain the causal effect because any indirect effects through empathy must go along the path $A_i \rightarrow M_i \rightarrow Y_i$. \citet[p. 207]{GreHaBul10} explain the logic of the falsification interpretation (see \citet[pp. 61--2]{CalDunTun24} for a similar description):
\begin{quote}
A more judicious approach at this juncture in the development of social science would be to encourage researchers to measure as many outcomes as possible when conducting  experiments\ldots With many mediators and only one intervention, this kind of experiment cannot identify which of the many causal pathways transmit the effect of the treatment, but if certain pathways are unaffected by the treatment,  one may begin to argue that they do not explain why [the intervention] works.
\end{quote}
In other words, under this argument, researchers view IOTs as a kind of falsification test---they hope they have the potential to rule out mechanisms such that passing the test has some informative value. Scholars often take this strategy explicitly because they are uncomfortable with the assumptions required to estimate the causal relationship between the mediator and the outcome. 

The establishment interpretation goes even further and interprets nonzero ATMs as suggestive evidence for a causal mechanism. Often, scholars test mediators that they assume or suspect have a particular causal effect on the outcome. For example, in our second empirical application on transitional justice and democratic attitudes, the authors interpret the significant effect of visiting the transitional justice museum as suggestive evidence of mechanism, operating under the assumption that emotions like fear of repressive regimes lead to a higher level of support for democracy \citep[][p.506]{balcells2022transitional}.

\subsection{The pitfalls of interpreting IOTs}

What do we actually learn when we perform IOTs? With a binary outcome, we know that an indirect effect must logically be between $-1$ and $1$ before we see any data. We can use the technique of nonparametric bounds to determine what values of an indirect effect are consistent with a set of empirical results and a set of maintained assumptions \citep{Manski95}. To do so, we use the linear programming approach of \citet{balk:pear:97} and \cite{sachs2023general}.\footnote{We ignore covariates in these derivations for presentational simplicity, but it is possible to incorporate covariates for either identification or for narrowing the bounds \citep{LevBonZen23}.}

Let $p_{ym \cdot a} = \mathbb{P}(Y_i=y, M_i=m \mid A_i=a)$ be the joint distribution of the outcome and the mediator within levels of treatment. In the context of our running example, this would be the joint distribution of prejudice and reactive empathy within levels of the perspective-getting treatment. \cite{imai2013experimental} derived the sharp nonparametric bounds for the ANIE under the mediation-skeptic assumptions (that is, Assumption~\ref{assm:randomization}) as
$$
\begin{aligned}
\max \left\{\begin{array}{l}-p_{10\cdot0}-p_{11\cdot0} \\ -p_{01\cdot1}-p_{11\cdot1}-p_{11\cdot0} \\ -p_{00\cdot1}-p_{10\cdot1}-p_{10\cdot0}\end{array}\right\} &\leqslant \bar{\delta}(0) \leqslant \min \left\{\begin{array}{l}p_{00\cdot0}+p_{01\cdot0} \\ p_{01\cdot1}+p_{11\cdot1}+p_{01\cdot0} \\ p_{00\cdot0}+p_{00\cdot1}+p_{10\cdot1}\end{array}\right\}, \\
\max \left\{\begin{array}{l}-p_{00\cdot1}-p_{01\cdot1} \\ -  p_{01\cdot1}-p_{01\cdot0}-p_{11\cdot0} \\ -p_{00\cdot0}-p_{00\cdot1}-p_{10\cdot0}\end{array}\right\} &\leqslant \bar{\delta}(1) \leqslant \min \left\{\begin{array}{l}p_{10\cdot1}+p_{11\cdot1} \\ p_{01\cdot    0}+p_{11\cdot0}+p_{11\cdot1} \\ p_{00\cdot0}+p_{10\cdot0}+p_{10\cdot1}\end{array}\right\}.
\end{aligned}
$$
We refer to these as the \emph{randomization bounds}. These bounds tell us the most we can learn about the ANIE from the joint distribution of the data (including the mediator) under the mediation-skeptical assumptions. Because the ATM is a deterministic function of the joint distribution, these bounds also tell us all that we can learn about the ANIE from an IOT. 

What can we learn from these sharp randomization bounds? Unfortunately, because $p_{ym\cdot a}$ is nonnegative, the randomization bounds will always contain positive values, negative values, and a precisely zero ANIE. This result implies that, under randomization alone, IOTs cannot support the falsification or establishment interpretations. In short, IOTs under randomization alone do not allow us to learn about the presence or sign of the ANIE. However, we do learn \emph{something} from the data. Logically, we know that the indirect effect must be between -1 and 1 for a binary outcome. The randomization bounds improve over those simple logical constraints and will rule out extreme values of the ANIE in either direction.

Why does randomization alone fail to justify the key IOT interpretations? There are two broad problems that IOTs face under the mediation-skeptical assumptions. First, we often lack credible causal evidence for the relationship between the mediator and the outcome, undermining the basis of the establishment interpretation. In the substantive example, scholars are apprehensive about interpreting any relationship between reactive empathy and prejudicial views as the causal effect of the former on the latter, even conditional on covariates. While it may be possible to rely on previous studies or experimental manipulation of $M_i$ in a separate arm to learn its average causal effect, we show below that knowledge of the average $M_i\rightarrow Y_i$ effect does not provide much information about the causal mechanisms.

The second and more fundamental problem with IOTs is that treatment heterogeneity between units renders average effects relatively uninformative for mediation quantities. For example, \cite{Glynn12} showed that the product of the average effect of $A_i$ on $M_i$ and the average effect of $M_i$ on $Y_i$ does not produce valid estimates of indirect effects when treatment effects can vary from unit to unit, undermining the idea that a zero ATM implies zero ANIE.\footnote{Indeed, \cite{GreHaBul10} recognize this when they say, ``this kind of analysis makes some important assumptions about homogeneous treatment effects, but the point is that this type of exploratory investigation may provide some useful clues to guide  further experimental investigation.'' We have not found researchers using IOTs to either acknowledge these assumptions or be as cautious in the interpretation of IOTs as \cite{GreHaBul10} recommend.} 
Returning to the prejudice application, we may estimate no effect of the perspective-getting intervention on the reactive empathy mediator, but this may mask positive and negative effects for equal-sized groups in the population, which average out to an overall null effect. In this setting, many values of indirect effect are possible because it may be the case that the group for whom perspective getting increases empathy has a positive effect of empathy on prejudice, which would lead to a positive indirect effect rather than the theorized negative indirect effect.

Treatment effect heterogeneity is pernicious because these types of unknown correlations undermine both the falsification and establishment interpretation. Detecting a significant nonzero average treatment effect on the mediator is uninformative because we learn nothing about the correlation between the individual-level effects of $A_i$ on $M_i$ and the individual-level effects of $M_i$ on $Y_i$. Below, we develop a formal framework to show how these two problems manifest and derive bounds that show precisely what we learn about indirect effects under various assumptions.

\subsection{Principal stratification approach to mediation}

One advantage of the binary setting we explore in this paper is that it allows us to follow a principal stratification approach \citep[see also,][]{BulGre21}. In particular, we can stratify all units into four groups based on how their mediator value responds to treatment:
\begin{itemize}
    \item mechanism compliers, $M_i(0) = 0, M_i(1) = 1$,
    \item mechanism defiers, $M_i(0) = 1, M_i(1) = 0$,
    \item always takers $M_i(0)= M_i(1) = 1$,
    \item never takers $M_i(0) = M_i(1) = 0$.
\end{itemize}
We borrow the terminology of compliers and defiers from the literature on instrumental variables to mean units that comply with or defy the proposed causal mechanism (rather than treatment assignment as in the instrumental variables setting). In the prejudice-reduction application, mechanism compliers would be the respondents for whom the perspective-getting intervention increases reactive empathy, whereas mechanism defiers are those for whom the intervention decreases reactive empathy.

We can define an individual-level indirect effect as 
$$
\delta_i(a) = Y_i(a, M_i(1)) - Y_i(a, M_i(0)),
$$
which makes it clear that the only individuals who will have an indirect effect are either mechanism compliers or defiers. That is $\delta_i(a) = 0$ for the always and never takers. This fact motivates the IOT approach: if no effect of $A_i$ on $M_i$ implies no indirect effect at the individual level, it feels natural that it should apply at the average level. Unfortunately, this is not the case.

To see why the IOT approach falls down without further assumptions, let us relate the average treatment effect on the mediator (ATM) to these principal strata. We can define the size of each of these groups as 
$$
\begin{aligned}
\rho_{st} \triangleq \text{Pr}[M_i(1)=s,M_i(0)=t].
\end{aligned}
$$
With these, we can show that ATM is equal to the proportion of mechanism compliers minus the proportion of mechanism defiers,
\begin{align*}
\alpha &= \E[M_i(1) - M_i(0)] = 1 \cdot \rho_{10} + (-1) \cdot \rho_{01} + 0 \cdot \rho_{11} + 0\cdot \rho_{00} = \rho_{10} - \rho_{01},
\end{align*}
where the second equality follows from an application of the iterated expectations. In other words, the ATM in the prejudice-reduction experiment would be the proportion of respondents whose reactive empathy increases because of treatment minus the proportion of respondents whose reactive empathy decreases because of treatment. The always-takers and never-takers contribute nothing because their values of $M_i$ never change with treatment. 

We can also write the ANIE in terms of these principal strata,
\begin{align}
\delta(a) &= \mathbb{E}[Y_i(a,M_i(1)) - Y_i(a,M_i(0))] \nonumber \\
&= \sum_{s=0}^{1} \sum_{t=0}^1 \mathbb{E}[Y_i(a,s) - Y_i(a,t) | M_i(1) = s, M_i(0) = t] \cdot \rho_{st} \nonumber \\
&= \underbrace{\mathbb{E}[Y_i(a,1) - Y_i(a,0) | M_i(1) = 1, M_i(0) = 0]}_{M\to Y \text{ effect for mechanism compliers}} \cdot  \rho_{10}\nonumber  \\
& \qquad -  \underbrace{\mathbb{E}[Y_i(a,1) - Y_i(a,0) | M_i(1) = 0, M_i(0) = 1]}_{M\to Y \text{ effect for mechanism defiers}} \cdot \rho_{01}. \label{eq:anie_no_mono}
\end{align}
The last expression shows that the ANIE is a function of the average effect of the mediator on the outcome for the mechanism compliers and defiers. The compliers contribute positive effects of the mediator while defiers contribute their negative effects. In the prejudice-reduction context, these are the causal effects of reactive empathy on prejudicial attitudes for those whose reactive empathy is positively (compliers) or negatively (defiers) affected by treatment. 

The IOT approach restricts itself from identifying the ATM, or $\rho_{10} - \rho_{01}$ in the principal strata framework. With no further assumptions, what can that tell us about mechanisms? From \eqref{eq:anie_no_mono}, we can see that we can have a nonzero ANIE even with an ATM of exactly zero. In particular, the ATM could be zero if there are equal numbers of mechanism compliers and defiers---that is, if an equal number of respondents' mediators are positively and negatively affected by treatment---but the effect of $M_i$ on $Y_i$ is different for these groups.  Thus, IOTs alone cannot rule out an indirect effect without further assumptions, leaving their interpretation as falsification tests suspect.  Furthermore, nonzero ATMs are also \emph{not sufficient} for indirect effects because any difference between $\rho_{10}$ and $\rho_{01}$ in \eqref{eq:anie_no_mono} can be offset by differences in the effect of $M_i$ on $Y_i$ in each of these groups. Thus, when we maintain the set of mediation-skeptical assumptions, intermediate outcomes provide relatively little information about causal mechanisms, at least in the form of indirect effects. 

The principal strata formulation of the ANIE in~\eqref{eq:anie_no_mono} can also clarify how the mediation assumptions identify the ANIE. One consequence of the mediator as-if randomization in Assumption~\ref{assm:m_randomization} is that the effects of treatment on the mediator and the effect of the mediator on the outcome are statistically independent. Thus, under this strong assumption, knowing that one is a mechanism complier or defier does not change the average effect of reactive empathy on prejudice. Omitting covariates for clarity, we can then simplify~\eqref{eq:anie_no_mono} as
\[
\begin{aligned}
\delta(a)
&= \mathbb{E}[Y_i(a,1) - Y_i(a,0) | M_i(1) = 1, M_i(0) = 0] \cdot  \rho_{10}\nonumber  \\
& \qquad -  \mathbb{E}[Y_i(a,1) - Y_i(a,0) | M_i(1) = 0, M_i(0) = 1] \cdot \rho_{01}. \nonumber \\
&= \mathbb{E}[Y_i(a,1) - Y_i(a,0)]\cdot \rho_{10} - \mathbb{E}[Y_i(a,1) - Y_i(a,0)]\cdot \rho_{01} \nonumber \\
&= \underbrace{\mathbb{E}[Y_i(a,1) - Y_i(a,0)]}_{M\to Y \text{ effect}} \cdot \underbrace{(\rho_{10} - \rho_{01})}_{A\to Y \text{ effect}}.
\end{aligned}
\]
Thus, the mediation assumptions imply powerful restrictions on effect heterogeneity, allowing us to use the product rule to identify the ANIE. Of course, the strength of this assumption is also why so many are skeptical of the assumption. 

None of these results depend on any uncertainty in our estimates. These results show what we could learn about indirect effects if we had access to infinite data and thus no sampling uncertainty. In reality, we will estimate these quantities, leading these tests to have even less informativeness.

\subsection{The hidden assumptions of intermediate outcome tests}

If intermediate outcomes are largely uninformative for mediation skeptics, why do we see them in widespread use? In this section, we show how to recover the intended informativeness of IOTs by making additional assumptions. We call these the \emph{hidden assumptions} since they are often implicitly invoked when using IOTs. 

We begin with assumptions on the relationship between the treatment and moderator since most researchers will be more comfortable with assumptions on the primary research design than on the effects of the mediator. We can recover part of the informativeness of the IOT approach by assuming that the treatment only affects the mediator in one direction. 
\begin{assumption}[Monotonic Mediator Response (MMR)]
\label{assm:pmono}
$\P[M_i(1) \geq M_i(0)] = 1.$
\end{assumption}
In words, Assumption~\ref{assm:pmono} states that an individual's potential mediator value is always weakly greater if they were to receive the treatment. This assumption rules out the existence of mediator-defiers in the population (that is, units for which $M_i(1)=0, M_i(0)=1$), meaning that $\rho_{01} = 0$ and the ATM will be equal to $\rho_{10}$. In the prejudice-reduction context, MMR requires that there be no respondent whose reactive empathy is lower because of treatment. \cite{KwoRot24} leverage this type of monotonicity assumption to develop a test of the sharp null that $M_i$ explains all of the treatment effect, $Y_i(a, m)=Y_i(m)$ for all $i$, $a$, and $m$. While this is an important test, we believe there are many causal effects that have many mechanisms (some of which are unmeasured), making the sharp null a less interesting test in these contexts.

Applying this assumption to \eqref{eq:anie_no_mono} allows us to write the ANIE as
\begin{align}
\delta(a) = \E[Y_i(a,1) - Y_i(a,0) | M_i(1) = 1, M_i(0) = 0] \cdot \rho_{10},\label{eq:anie_mono_1}
\end{align}
i.e., a scaled (by $\rho_{10}$) controlled direct effect (CDE) of $M_i$ on $Y_i$ among the mechanism compliers. Thus, we recover a weak ``product rule'' interpretation of the indirect effect under MMR. We can write the ANIE as the product of the effect of the treatment on the mediator and the effect of the mediator on the outcome among units whose mediator is affected by treatment, or 
\begin{align}
\delta(a)
&= \mathbb{E}[Y(a,1) - Y(a,0) | M(1) = 1, M(0) = 0] \cdot  \text{ATM}.\label{eq:anie_mono_2}
\end{align}
Thus, under MMR, an average effect of treatment on the moderator of zero rules out an average indirect effect, and the IOT can now properly function as a falsification test. Unfortunately, intermediate outcome tests remain insufficient for establishing causal mediation because even if the ATM is nonzero, the effect of $M_i$ on $Y_i$ for the mechanism compliers may still be zero. 

Is the MMR assumption plausible? The answer to this question will, of course, depend on the context. There may be many settings where this is a perfectly reasonable assumption. A survey experiment with a prompt designed to anger respondents is unlikely to make some respondents less angry. The difficulty with this assumption is that it must be true for all units to make the above interpretations correct, which may be difficult to sustain in many cases. Below, we discuss one way to test the assumption.

\subsection{Sharp bounds for the indirect effect}\label{subsec:sharp}

The previous section showed how to use IOTs to rule out causal mechanisms when the effect of the treatment on the mediator is precisely zero. What about situations where the ATM is small but nonzero? We now derive sharp bounds for the indirect effect under different assumptions. The first result imposes MMR and provides bounds in terms of the ATM. 

\begin{proposition}\label{prop:mmr}
Under Assumption~\ref{assm:pmono}, the sharp bounds for the ANIE are 
\begin{align}
\max\left\{-ATM, -p_{10\cdot 0}\right\} \leq &\; \delta(0)  \leq \min\left\{ATM,p_{00\cdot 0} \right\}, \\
\max\left\{-ATM, -p_{01\cdot 1}\right\} \leq &\; \delta(1)  \leq \min\left\{ATM,p_{11\cdot 1} \right\}
\end{align}
\end{proposition}

These sharp bounds under the MMR assumption provide a more nuanced interpretation of IOTs. When the ATM is relatively small (compared to $p_{10\cdot 0}$, $p_{00\cdot 0}$, $p_{01\cdot 1}$, and $p_{11\cdot 1}$), the sharp bounds for the ANIE will be $[-\text{ATM}, \text{ATM}]$. Thus, when the ATM is close to 0, the bounds will also be close to zero, and we will be able to rule out large indirect effects. At the extreme, if the ATM is zero, then the ANIE is also point identified as zero, as the discussion above implies. This result implies that IOTs under MMR are ``approximately necessary'' in that small (zero) ATMs imply small (zero) indirect effects. Again, however, we emphasize that this interpretation of an IOT relies on the monotonic mediator response assumption, which cannot be guaranteed to hold by design.

One potential justification for the IOT approach is if past empirical evidence or theory implies there should be an average effect of the mediator on the outcome. If the treatment ``activates'' the pathway from $A_i$ to $M_i$ in the experimental data (that is, there is an ATM), then this should provide sufficient evidence to establish the causal mechanism if the supposition is true, or so the reasoning goes. In the next proposition, we provide sharp bounds on the indirect effect under MMR and knowledge that the average treatment effect of $M_i$ on $Y_i$ is positive. Even under these very strong assumptions, the bounds always include an indirect effect of zero. 

\begin{proposition}\label{prop:ym_ate}
Suppose that Assumption~\ref{assm:pmono} holds and $\E[Y_i(1, 1) - Y_i(1, 0)] \geq 0$. Then, 
\begin{equation}
\max \left\{\begin{array}{l} -ATM \\ p_{10\cdot 1} -p_{10\cdot 0}-p_{00\cdot0} \\ -p_{11\cdot1}-p_{00\cdot1}-p_{10\cdot0} \\ -p_{01\cdot 1}\end{array}\right\} \leq \bar{\delta}(1) \leq \min \left\{\begin{array}{l}ATM \\ p_{11\cdot1}+p_{10\cdot 0}+p_{00\cdot0} \\ 2p_{11\cdot 1}+p_{00\cdot1}+p_{00\cdot 1} \\ p_{11\cdot 1} \end{array}\right\}, \\
\end{equation}
and these bounds are sharp. Furthermore, all lower bounds are nonpositive, and all upper bounds are nonnegative. 
\end{proposition}

Proposition~\ref{prop:ym_ate} shows that even if we assume monotonic mediator response \emph{and} a positive average treatment effect of the mediator on the outcome, we still cannot learn the direction of the indirect effect. Equation~\eqref{eq:anie_mono_2} shows why this is the case---the indirect effect focuses on the effect of $M_i$ on $Y_i$ for the mechanism compliers, whereas the average treatment effect of $M_i$ on $Y_i$ is the effect averaged across all units. Thus, IOTs cannot be sufficient tests for causal mechanisms even under these assumptions. 

%% Add propositions about M->Y here
Lastly, is there a scenario under which IOTs do constitute sufficient evidence for causal mechanisms? In Supplemental Materials~\ref{sm:full_mono}, we find that a full monotonicity assumption---namely, that all causal effects are in a single direction at the individual level---can allow us to establish the ANIE when the IOTs are nonzero. Of course, to make these restrictions, we would have to know the effect of the mediator on the outcome, which is often unrealistic.

Given the importance of the MMR assumption, how can we determine whether it holds? Testing MMR is challenging because it restricts individual-level treatment effects, which are unidentified. We can, however, exploit what MMR implies for other quantities. In particular,  MMR states that individual's treatment effect of $A_i$ on $M_i$ ($M_i(1)-M_i(0)$) be greater than zero, which implies that any conditional average treatment effects (CATEs) of the treatment on the mediator must be positive or equal to zero (i.e., $\mathbb{E}[M_i(1)-M_i(0)|X_i=x] \geq 0$). CATEs capture the average treatment effect of $A_i$ on $M_i$ for different subgroups of the data, defined by covariates $X_i$. Fortunately, the CATEs, unlike the individual-level treatment effects, are identified under some restrictions, allowing us to conduct a falsification test. We can use modern machine learning methods such as causal forests \citep{wager2018estimation} to estimate the distribution of CATEs for a given set of covariates. If any of the estimates are below zero, this indicates a violation of the MMR assumption. Of course, even if all CATE estimates are above zero, this does not rule out all possible violations of MMR: there may still be individuals within each subgroup with negative individual effects that are canceled out by those with positive effects.

What if we cannot sustain the MMR assumption but wish to impose some structure beyond randomization on our problem? In Supplemental Materials~\ref{sm:limited_defiers}, we derive the sharp bounds for the ANIE under an assumed upper bound on the proportion of mechanism defiers. These bounds would allow researchers to conduct a sensitivity analysis to see how their inferences depend on this quantity.

\subsection{Estimation and inference for the bounds}

The above bounds are all functions of the population probabilities $p_{ym\cdot a}$, but researchers only have access to the observed data $\{(A_i, M_i, Y_i, \bX_i)\}_{i=1, \ldots, n}$. To perform estimation and inference for these bounds, we rely on the intersection bounds approach of \cite{CheLeeRos13}. Traditional plug-in estimators for the bounds tend to be biased so that the plug-in bounds are narrower than the population bounds because of the use of the minimum and maximum operators. Furthermore, the estimated standard error of the selected bounding expression would ignore the uncertainty that the other bounding expressions also have, which contributes to uncertainty about the overall bound. The approach of \cite{CheLeeRos13} uses precision-corrected versions of each bounding expression to obtain (a) upper and lower bound estimates that are half-median unbiased and (b) provides confidence intervals for the true indirect effect that cover the true value of the indirect effect at nominal rates. Half-median unbiased means that the probability that the estimated upper (lower) bound is greater (less) than the population upper (lower) bound is at least one-half. This is a common desirable property of estimators in partially identified settings like ours where previous work has shown that traditional unbiasedness is not possible to achieve \citep{HirPor12}. We implement these empirical procedures in the next section for our two motivating examples.

\section{Empirical Analyses}\label{sec:applications}
 
 \subsection{Reducing Outgroup Prejudice}

We first re-analyze data from the \citet{kalla2023narrative}  study of intergroup prejudice.  In one of their experimental studies, the authors showed respondents a picture of an outgroup member (either undocumented immigrants or transgender people) and then asked them to engage in a randomly assigned exercise corresponding to different narrative strategies.\footnote{These are traditional perspective-taking, analogic perspective-taking, perspective getting with an essay task to summarize the story heard, and perspective getting without the essay task.} We focus on the pooled (i.e., across the unauthorized immigrants and transgender people arms) effects of the ``perspective getting with essay'' treatment relative to the baseline condition since this treatment yields the highest effect on prejudice reduction of the four treatment conditions considered. In this treatment condition, respondents read a story about an outgroup member and were then asked to summarize the story they just read. Overall, the authors find that the treatment significantly reduces respondents' prejudice against outgroup members.

To assess which mechanisms explain these effects, the authors asked respondents four additional questions as potential mediators: (i) whether respondents have a lot in common with the outgroup (self-outgroup merging); (ii) whether the suffering of the outgroup is of concern (reactive empathy); (iii) whether outgroup members face challenges that are no fault of their own (attributional thinking), and (iv) whether the outgroup's name elicits thoughts about individual people or a group of people (entativity). The authors implement four separate IOTs on these four mediators. 

\begin{figure}[t]
\vspace{1cm}
\begin{center}
\includegraphics[width=1.1\textwidth]{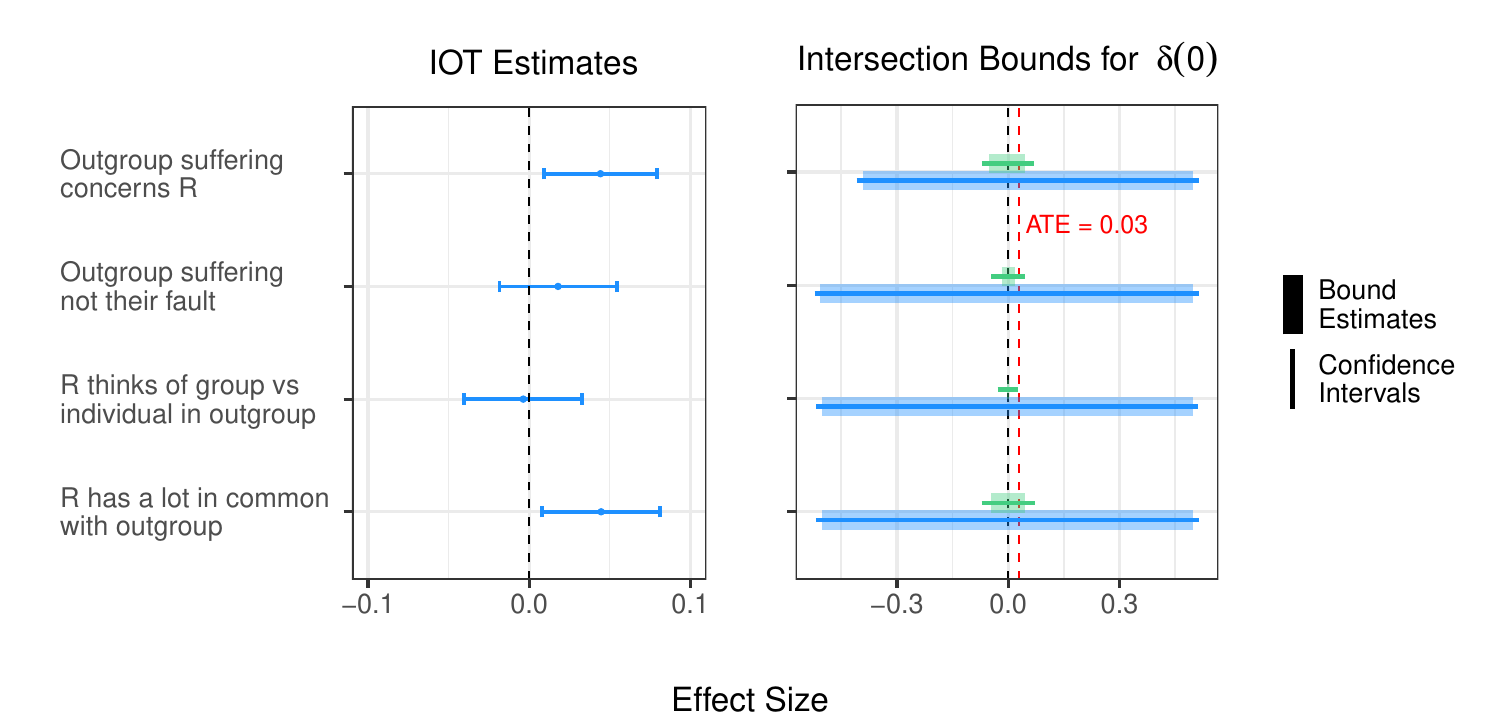}
\caption{Establishing causal mechanisms for the effect of perspective-getting on prejudice reduction \citep{kalla2023narrative}. The figure displays IOT point estimates (left panel) and sharp bounds estimates (thick, transparent lines) for the indirect effect of perspective-getting with an essay assignment on prejudice reduction via each of the putative mediators at reference level $A=0$ ($\delta(0)$), as well as the 95\% confidence intervals for these bounds (thin lines). Randomization bounds are shown in blue, while MMR bounds derived under Assumption \ref{assm:pmono}) are shown in green. The dotted black line marks zero, while the dotted red line marks the ATE (0.03).}
\label{fig: kalla}
\end{center}
\end{figure}

Since our sharp bounds for the indirect effect (as described in Section \ref{subsec:sharp}) rely on dichotomous mediators and outcomes, we first dichotomize the four mediators and the outcome variable at their medians. We then implement the IOT by estimating ATM effects for each mediator.\footnote{We opt not to use covariate adjustment for these ATM effects, which is valid given the random treatment assignment.} As shown in the right panel of Figure \ref{fig: kalla}, these IOT estimates largely replicate the direction, magnitude, and statistical significance of those reported by \cite{kalla2023narrative}: perspective-getting with an essay task has a positive and statistically significant effect on whether the suffering of the outgroup is of concern (0.04, 95\% confidence interval of [0.03, 0.06]) and whether respondents have a lot in common with the outgroup (also 0.04, 95\% confidence interval of [0.03, 0.06]).\footnote{Note that the authors also find a positive effect of perspective-getting with an essay task on whether outgroup members face challenges that are no fault of their own, a finding that does not replicate without controlling for the battery of covariates included in the authors' models after dichotomization.}

In their original analysis, \cite{kalla2023narrative} provide a nuanced interpretation of the potential mechanisms. For example, the experiment contains two placebo treatments with null effects on the main outcomes (that is, null estimated ATEs) but significant effects on two potential mediators (attributional thinking and self-outgroup merging). The authors argue that this combination implies that these two mediators are unlikely to explain the ATE of the primary perspective-getting treatment. Combined with the IOT evidence on concern about outgroup suffering, the authors note that ``perspective-getting reduces prejudice by prompting empathy towards that group or through other unmeasured mediators'' and ultimately conclude that unmeasured mediators may be the more important of these two. Thus, the authors lean more toward the falsification interpretation of IOTs rather than the establishment interpretation but ultimately cast doubt on how much can be learned from the IOT analysis.

What conclusions can we reach with our nonparametric bounding analysis? The right panel of Figure~\ref{fig: kalla} displays the sharp bounds for the indirect effect with the treatment reference category of $A=0$, i.e., $\delta(0)$, and calibrates these bounds both against zero (dotted black line) and the overall ATE of the intervention (dotted red line), which was $0.03$. 

For both the randomization bounds (wider blue-colored bounds) and the MMR bounds (narrower green-colored bounds), we present the sharp bounds (thick, transparent lines) in addition to their 95\% confidence intervals (thin lines). The estimated randomization bounds are extremely wide and approximately center around $0$, providing little to no information about the existence of an indirect effect for each of the mediators considered, nor their direction or magnitude. In particular, the randomization bounds estimated for reactive empathy---a potentially important mediator based on the IOT test---is $[-0.39, 0.50]$, an interval that contains a large range of positive and negative effect sizes, contains zero, and includes a set of indirect effect sizes on the order of 10 times larger in absolute value than the estimate recovered by the IOT approach. In short, the randomization bounds do not allow us to conclude that any of the four mediators considered mediate the effect of perspective-getting on prejudice reduction, even those mediators that ``pass'' an IOT (i.e., those mediators that are themselves affected by the treatment), such as reactive empathy and attributional thinking. 

Does assuming mediator monotonicity -- that the treatment affects the mediator in one direction for \textit{all} respondents -- enable us to recover the informativeness of the IOT approach? The green bands in Figure \ref{fig: kalla} panel B address this question. Two points are of particular note. First, the bounds shrink substantially compared with those computed without restriction on the distribution of potential outcomes. For example, the bounds estimate for reactive empathy shrinks from $[-0.39, 0.50]$ under no assumption to $[-0.05,  0.043]$ under MMR, a considerably smaller interval that ranges from approximately $-ATM$ to $+ATM$. 

Second, and despite this, for the two putative mediators with nonzero IOT estimates (concern about outgroup suffering and feeling commonality with the outgroup), the bounds are uninformative about either the existence or the sign of the relevant indirect effect. The bounds all contain zero and a large range of positive and negative values, including the overall ATE (red dotted line). In other words, the observed data are consistent with the indirect effect via these mediators being zero, as well as with the indirect effect via these mediators being equal to the ATE (and thus, consistent with these mediators serving as the \emph{sole} mechanism through which the ATE operates). Only for the group vs. individual mediator -- and to a lesser extent, the ``outgroup suffering is not their fault'' mediator -- is the MMR bounds estimate extremely narrow: in this case, it almost collapses on zero, a result which mirrors the IOT finding of a minimal effect of treatment on those mediators. This result highlights the ``approximately necessary'' condition of IOTs under MMR, in that small ATMs imply small indirect effects.

\begin{figure}[t]
\vspace{1cm}
\begin{center}
\includegraphics[width=0.9\textwidth]{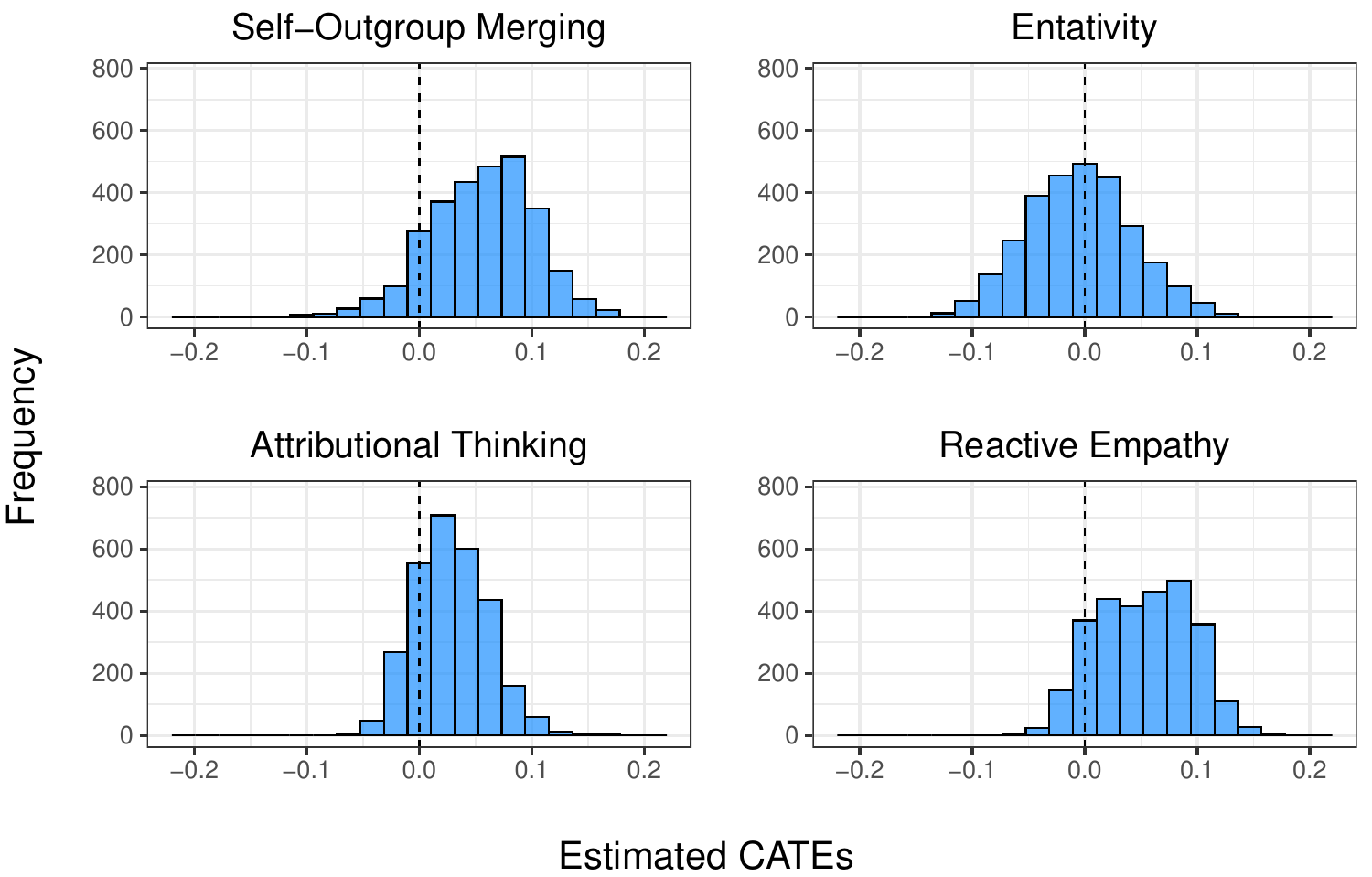}
\caption{Validating the MMR Assumption for the effect of perspective-getting on prejudice reduction \citep{kalla2023narrative}. The figure displays a histogram for estimated CATE values for the effect of $A_i$ on $M_i$ for each of the four putative mediators. CATEs are estimated using causal forests \citep{wager2018estimation} with 2000 trees.}
\label{fig:kalla_cate}
\end{center}
\end{figure}

Is the MMR a reasonable assumption in this setting? At a theoretical level, for the group vs. individual mediator proposed by \cite{kalla2023narrative}, this means assuming that no individual for whom hearing about the feelings of prejudice of trans people or immigrants would make them more likely to view them as a group rather than as individuals. Given the amorphous nature of this kind of evaluation, we think it is difficult to maintain this strong assumption in the absence of additional evidence. Empirically, we can implement our falsification test of MMR, as discussed in Section ~\ref{subsec:sharp}. Specifically, we estimate CATEs ($\mathbb{E}[M_i(1)-M_i(0)|X_i=x]$) using causal forests. This flexible machine-learning method facilitates the discovery of heterogeneous treatment effects that may be highly non-linear across covariate values \citep{wager2018estimation}. This nonparametric approach is particularly advantageous in our context, where our falsification goal is to see whether any CATE falls below zero---capturing CATEs flexibly without imposing any particular parametric functional form allows us to assess this claim honestly. Figure~\ref{fig:kalla_cate} plots histograms for each of the four putative mediators. We see that in all cases, a non-trivial proportion of estimated CATEs falls below zero. In this way, MMR appears to fail for all mediators in this empirical setting. Thus, in this particular case, we believe the uninformative randomization bounds are more appropriate, which is in line with the authors' inconclusive interpretation of mechanisms in the original paper. 

\subsection{Transitional Justice Museums and Support for Democracy}

We now turn to the field experiment of \cite{balcells2022transitional} on transitional justice museums. Around 500 university students in Santiago, Chile, were randomly assigned to either a treatment or control condition. Students in the control group were issued an end-line survey with questions about their political attitudes; students in the treatment group participated in a one-hour tour of the Museum of Memory and Human Rights and then immediately after completed a survey analogous to that administered to the control group. The authors find that museum visitation significantly increased students' support for democratic institutions and transitional justice policies and decreased support for repressive institutions.

\begin{figure}[t]
\vspace{1cm}
\begin{center}
\includegraphics[width=1.1\textwidth]{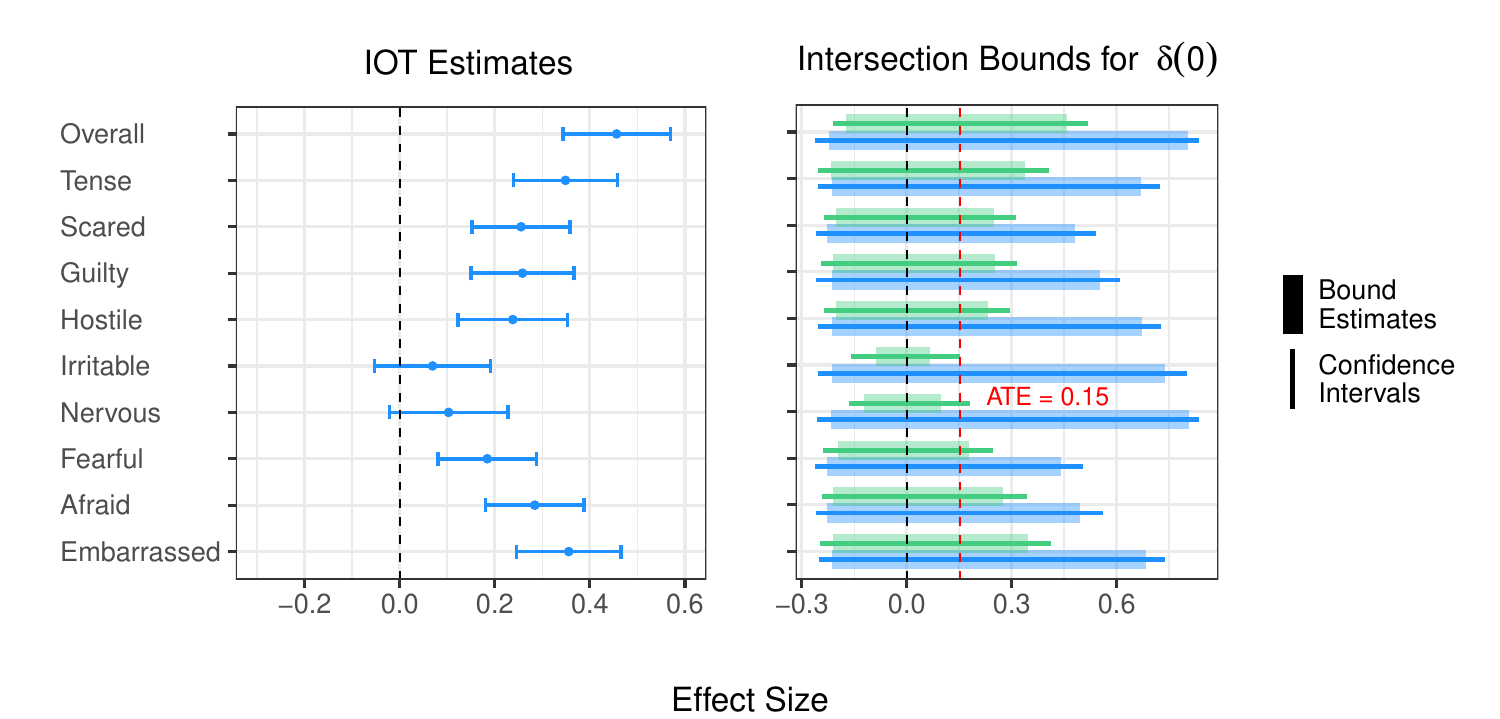}
\caption{Establishing causal mechanisms for the effect of a transitional justice museum visit on support for democratic institutions \citep{balcells2022transitional}. The figure displays IOT point estimates (left panel) and sharp bounds estimates (thick, transparent lines) for the indirect effect of the museum intervention via each of the putative mediators at reference level $A=0$ ($\delta(0)$), as well as the 95\% confidence intervals for these bounds (thin lines). Randomization bounds are shown in blue, while MMR bounds derived under Assumption \ref{assm:pmono}) are shown in green. The dotted black line indicates zero; the dotted red line indicates the ATE (0.15) of the museum visit on support for democracy.}
\label{fig:balcells}
\end{center}
\end{figure}

The authors posit that museums’ narratives of past events may have elicited emotional reactions among their visitors. To test this hypothesis, they measured treated and untreated students' responses to the Positive and Negative Affect Schedule, which asks respondents to consider a range of positive and negative emotions and to indicate which they feel in the present moment \citep{watson1988development}. We focus on the negative emotion questions as the authors find treatment effects on these emotions to be substantively larger and more consistent than the positive emotion questions. Similar to the previous example, we dichotomize all mediators as well as the outcome variable at their medians. The left panel of Figure~\ref{fig:balcells} reports the IOT estimates, which replicate the patterns reported by \cite{balcells2022transitional}: the visit to the transitional justice museum has a positive and statistically significant effect on the overall negative emotions experienced by respondents (0.46, 95\% confidence interval of $[0.35,0.57]$). In particular, the authors find positive and significant treatment effects on feeling tense, scared, guilty, hostile, fearful, afraid, and embarrassed. Drawing on the establishment interpretation of IOTs, \cite{balcells2022transitional} argue that these findings indicate that the emotional experience of a transitional justice museum may act as a key mechanism by which the museum visit alters political attitudes.

While the IOTs could indicate that the strong and significant emotional reactions elicited by a museum visit could serve as an important mediating pathway for the overall museum effect, the nonparametric bounds in the right panel of Figure~\ref{fig:balcells} provide a more inconclusive interpretation. Paralleling the previous canvassing example, we make three remarks about the bounds on $\delta(0)$. First, all randomization bounds are extremely wide, and all contain zero. For example, the bounds for the overall negative emotion outcome is $[-0.22, 0.80]$, which confirms our proposition that randomization alone can neither establish nor rule out the existence of mechanisms. Second, the MMR bounds are significantly shorter than their randomization counterparts, although they still contain zero. Therefore, while the MMR assumption seems relatively plausible under this experimental setting---that compared to the control condition, visiting a museum recounting the country's past political violence only increases negative emotions such as fear and guilt---the bounds derived under this assumption remain uninformative about mediation via each of the variables considered. Moreover, as we show in the Supplemental Materials (Figure~\ref{fig:balcells_cate}), the estimated CATEs from a causal forest analysis for each of the individual emotions do not fall below zero. Finally, smaller IOT coefficients correspond to shorter bounds under the MMR assumption. For instance, the effect of the museum visit on ``feeling irritable'' has a point estimate of 0.07, the smallest among all estimates. Correspondingly, the bound for ``feeling irritable'' is also the shortest, ranging from $[-0.09, 0.07]$, which is approximately $-ATM$ to $+ATM$. Even for these mediators, however, the confidence intervals are close to the overall ATE, meaning we cannot rule out the possibility that they account for the entire effect of the museum intervention.

\section{What Is the Point of Mechanism Tests?}\label{sec:why}

As these examples show, the question of mechanisms has become an important part of empirical political science. Indeed, according to our literature review, over a third of recent empirical papers in top political science journals deploy some form of mechanism testing. Why is there an emphasis on this project if many believe the assumptions necessary to sustain such tests are implausible? 

In the social sciences, the idealized version of empirical research happens by a process where researchers (a) develop a relatively broad causal theory (or explanation), (b) use this theory to derive specific hypotheses about specific causal effects, and (c) use convincing causal identification and suitable data and estimators to test these hypotheses.  One hopes the connection between a hypothesis and the estimated causal effect will be strong, but one hypothesis (and the resulting causal effect) may be consistent with several broader causal theories. Researchers hope to use mechanism tests to bridge this gap by showing the path by which the treatment affects the outcome. A causal mechanism helps generalize a specific causal effect into a more general form, allowing us to hopefully better understand all situations in which this mechanism is activated. Unfortunately, there are many instances where all approaches to establishing indirect effects, including IOTs, rely on shaky assumptions that are difficult to justify. What should researchers do in that case?

We encourage researchers to pursue the time-tested approach of ``implication analysis'' \citep{LieHor08} or ``inference to the best explanation'' \citep{SpiSte24}, whereby researchers generate many different hypotheses that differentiate between the proposed theory and alternatives. The benefit of this approach is that these secondary hypotheses can be completely unrelated to the direct or indirect effects of the main causal effect of interest. This approach can free researchers from focusing on research designs with assumptions ill-suited to a particular setting. Of course, as we have articulated here, we must be clear about the assumptions needed to draw conclusions from these tests. In many ways, the main goal of \cite{kalla2023narrative} takes this exact approach. Earlier work showed that interpersonal conversations can lead to reductions in exclusionary attitudes, but it was not clear \emph{why} these conversations worked. The experimental design of \cite{kalla2023narrative} is tailored to test different types of narrative strategies to help adjudicate between different theoretical mechanisms of different narrative strategies. Thus, their primary analysis helps clarify causal mechanisms without resorting to IOTs, effect heterogeneity, or causal mediation.

\section{Conclusion}\label{sec:conclusion}

This paper shows how a popular way of assessing causal mechanisms---the intermediate outcome test---relies on strong and often undisclosed assumptions. In these tests, authors present the average effects of treatment on the mediator and often argue that these effects show plausible causal pathways of the overall effect of the treatment on the outcome. However, our analysis shows that IOTs can neither establish nor rule out indirect effects through a mediator under randomization of treatment alone. We showed that IOTs \textit{can} rule out indirect effects if the effect of treatment on the mediator can only move units in one direction, which is a very strong monotonicity assumption. Furthermore, IOTs cannot establish an indirect effect even if we assume a positive average effect of the mediator on the outcome. We derive sharp nonparametric bounds for the indirect effect under these various assumptions and estimate these bounds in two empirical applications. We also introduced a machine learning framework for evaluating the crucial monotonicity assumption. Finally, we also discussed how researchers can assess causal mechanisms when the assumptions needed to make inferences about indirect effects are questionable. 

We believe there remains much to study in the field of causal mechanisms.  In particular, future research should attempt to unify different accounts of ``mechanisms'' in a single broad framework. Currently, mechanisms are sometimes used interchangeably with ``indirect effects,'' but other times, scholars use mechanisms to refer to a theoretical construct that can explain an effect without reference to any mediation at all. In epidemiology, there is a tradition of defining causal mechanisms in the ``sufficient cause framework,'' where authors seek to find sets of variable levels that are sufficient to produce a particular outcome \citep{VanderWeele09a}. Connecting the ideas of mediation and mechanism to the theoretical underpinnings of the social sciences would help scholars develop better empirical tests of their theories.

\if1\submission{}
\begin{spacing}{1.5}
\fi
\newpage
\bibliographystyle{apsr}
\bibliography{ref.bib}

\if1\submission{}
\end{spacing}
\fi

\if0\submission{}
\clearpage
\begin{appendices}
\renewcommand{\thesection}{\Alph{section}}
\renewcommand\thefigure{SM.\arabic{figure}}
\renewcommand\thetable{SM.\arabic{table}}
\setcounter{lemma}{0}
\renewcommand{\thelemma}{SM.\arabic{lemma}}

\input{smuggling_appendix.tex}
\end{appendices}
\fi

\end{document}

%% file: smuggling_appendix.tex
\section{Overview of the linear programming approach}

The bounds we have derived in the main text come from an optimization problem that can be solved via a linear programming approach. This approach largely follows the discussion of \cite{imai2013experimental} with different maintained assumptions. For each indirect effect, one of the values in the causal contrast is identified. For instance, with $\overline{\delta}(1)$, we only need to consider $\E[Y_i(1, M_i(0))]$ since $\E[Y_i(1, M_i(1))] = \E[Y_i(1)] = \E[Y_i\mid A_i = 1]$ is identified without further assumptions. The goal of the partial identification, then, is to find the maximum and minimum value of $\E[Y_i(1, M_i(0))]$ that is consistent with the data and the maintained assumptions. To do this, we introduce an augmented set of principal strata,
$$
\psi_{y_1y_0m_1m_0} = \E[Y_i(1, 1) = y_1, Y_i(1, 0) = y_0, M_i(1) = m_1, M_i(0) = m_0].
$$
Note that we only need to consider the potential outcomes under treatment $Y_i(1, m)$ here since they are the only quantities that appear in the ANIE.

We can write the subject of optimization, the cross-world counterfactual, in terms of the principal strata as
\begin{equation}\label{eq:cross_world}
\E[Y_i(1, M_i(0))] = \sum_{y = 0}^1\sum_{m=0}^1 \psi_{1ym1} + \psi_{y1m0}    
\end{equation}
There are several constraints that must hold on the principal strata dictated by the axioms of probability or assumptions about how the observed data and potential outcomes relate. First, we have the logical constraint
$$
\sum_{y_1 = 0}^1 \sum_{y_0=0}^1 \sum_{m1=0}^1\sum_{m0=0}^1 \psi_{y_1y_0m_1m_0} = 1, 
$$
and the relationship between the observed probabilities and the principal strata,
$$
\begin{aligned}
p_{ym\cdot 1} = \begin{cases} \sum_{y_0 = 0}^1 \sum_{m_0=0}^1 \psi_{yy_0mm_0} & \text{if } m = 1 \\  \sum_{y_1 = 0}^1 \sum_{m_0=0}^1 \psi_{y_1ymm_0} & \text{if } m = 0  \end{cases}.
\end{aligned}
$$
These two restrictions (along with a nonnegativity constraint on all $\psi$ values) are the only restrictions from just randomization. Assumption~\ref{assm:pmono} (MMR) implies that any mechanism defier strata has zero probability of occurring,
$$
\psi_{y_1y_001} = 0 \qquad \forall y_1 \in \{0,1\}, y_0 \in \{0,1\}. 
$$
Finally, for Proposition~\ref{prop:ym_ate}, we can add the restriction that the effect of $\E[Y_i(1, 1) - Y_i(1, 0)] \geq 0$ with
$$
\sum_{m_1 =0} ^1 \sum_{m_0=0} ^1\psi_{10m_1m_0}  - \sum_{m_1 =0} ^1 \sum_{m_0=0} ^1\psi_{01m_1m_0} \geq 0.
$$

Using standard linear programming techniques, we can obtain bounds by maximizing or minimizing \eqref{eq:cross_world}  with respect to these restrictions. We can then combine these with the already identified expressions for $\E[Y_i(1, M_i(1))]$ to obtain sharp bounds for $\overline{\delta}(1)$. A similar approach can obtain bounds for $\overline{\delta}(0)$. 

\newpage
\section{Bounds under Limited Defiers}\label{sm:limited_defiers}

The MMR assumption may be too strong in many cases, so we now derive bounds under a weaker restriction on the mechanism defiers. In particular, we assume the proportion of defiers is bound
\begin{equation}\label{eq:limited_defiers}
\rho_{01} = \sum_{y_1=0}^1 \sum_{y_0 = 0}^1 \psi_{y_1y_001} \leq \gamma. 
\end{equation}
This restriction can easily be incorporated into the linear programming approach described above.

\begin{proposition}\label{prop:limited_defiers}
Suppose that Assumption~\ref{assm:randomization} and restriction~\eqref{eq:limited_defiers} holds. Then, we have
\begin{equation}
\max \left\{\begin{array}{l} 
-ATM - 2\gamma \\ 
-p_{01\cdot 1} - p_{00\cdot 1} \\
-p_{00 \cdot 1} - p_{10\cdot 0} - p_{00\cdot 0} \\
-ATM - p_{00\cdot 1} - \gamma \\
-p_{01 \cdot 1} - p_{11\cdot 0} - p_{01 \cdot 0} \\
-p_{01\cdot 1} - \gamma
\end{array}\right\} \leq \bar{\delta}(1) \leq \min 
\left\{\begin{array}{l}
ATM + 2\gamma \\ 
p_{11\cdot 1}+p_{10\cdot 1} \\ 
p_{11\cdot 1}+\gamma \\ 
p_{11\cdot 1} +p_{11\cdot 0} +p_{01\cdot 0} \\
ATM + (1 - p_{11 \cdot 1}) \\
ATM + p_{10\cdot 1} + \gamma 
\end{array}\right\}. \\
\end{equation}
\end{proposition}

We derive these bounds under full generality about the relative size of $\gamma$ to observed data distribution, but we can provide more interpretable bounds if we assume that the proportion compliers is low relative to the obsered strata probabilities. In particular, if $\gamma \leq \min_{y,m} p_{ym\cdot 1}$, then we have 
\begin{equation}
\max\{-ATM - 2\gamma, -p_{01\cdot 1} - \gamma\} \leq \bar{\delta(1)} \leq \min\{ATM + 2\gamma, p_{11\cdot 1} + \gamma\}.
\end{equation}
One takeaway from these bounds is that when the ATM is zero (and the defiers group is small), the ANIE will be bounded between $-2\gamma$ and $2\gamma$. Substantively, this means that when there is no average effect of the treatment on the mediator, we can rule out indirect effects that are more twice the hypothesized size of the defier group. In other words, if the defier group is allowed to be half the size of the ATE, then a zero ATM would still be consistent with the possibility of an ANIE that explains the entire treatment effect. 

\newpage
\section{ANIE Bounds under Full Monotonicity}\label{sm:full_mono}

In this section, we formalize the full monotonicity assumption and derive the corresponding bounds on ANIE. We start by stating the assumption as follows:

\begin{assumption}[Full Monotonicity]
\label{assm:fmono}
$$\P[M_i(1) \geq M_i(0)] = 1, \P[Y_i(1,m) \geq Y_i(0,m)] = 1, \P[Y_i(a,1) \geq Y_i(a,0)] = 1$$
\end{assumption}

Notice that full monotonicity assumption is strictly stronger than the MMR assumption. Moreover, we derive the ANIE bounds using the same linear programming procedure as described above:

\begin{proposition}
\label{prop:fmono}
Under Assumption~\ref{assm:fmono}, the sharp bounds for the ANIE are\footnote{With straightforward algebraic manipulations, these bounds can be shown to be equivalent to those derived by \cite{cai2008bounds} using Balke's algorithm.} 

$$
\begin{array}{l}\max \left\{\begin{array}{c}0 \\ p_{10 \cdot 1}-p_{10 \cdot 0}\end{array}\right\} \leqslant \delta(0) \leqslant p_{10 \cdot 1}+p_{11 \cdot 1}-p_{10 \cdot 0} \\ \max \left\{\begin{array}{c}0 \\ p_{01 \cdot 0}-p_{01 \cdot 1}\end{array}\right\} \leqslant \delta(1) \leqslant p_{00 \cdot 0}+p_{01 \cdot 0}-p_{01 \cdot 1},\end{array}
$$
\end{proposition}
This result shows that, under the full monotonicity assumption, the bounds on ANIE no longer cross zero.

\newpage
\section{Additional MMR Falsification Tests}
\label{sm:mmr_falsification}

\begin{figure}[h]
\vspace{1cm}
\begin{center}
\includegraphics[width=0.9\textwidth]{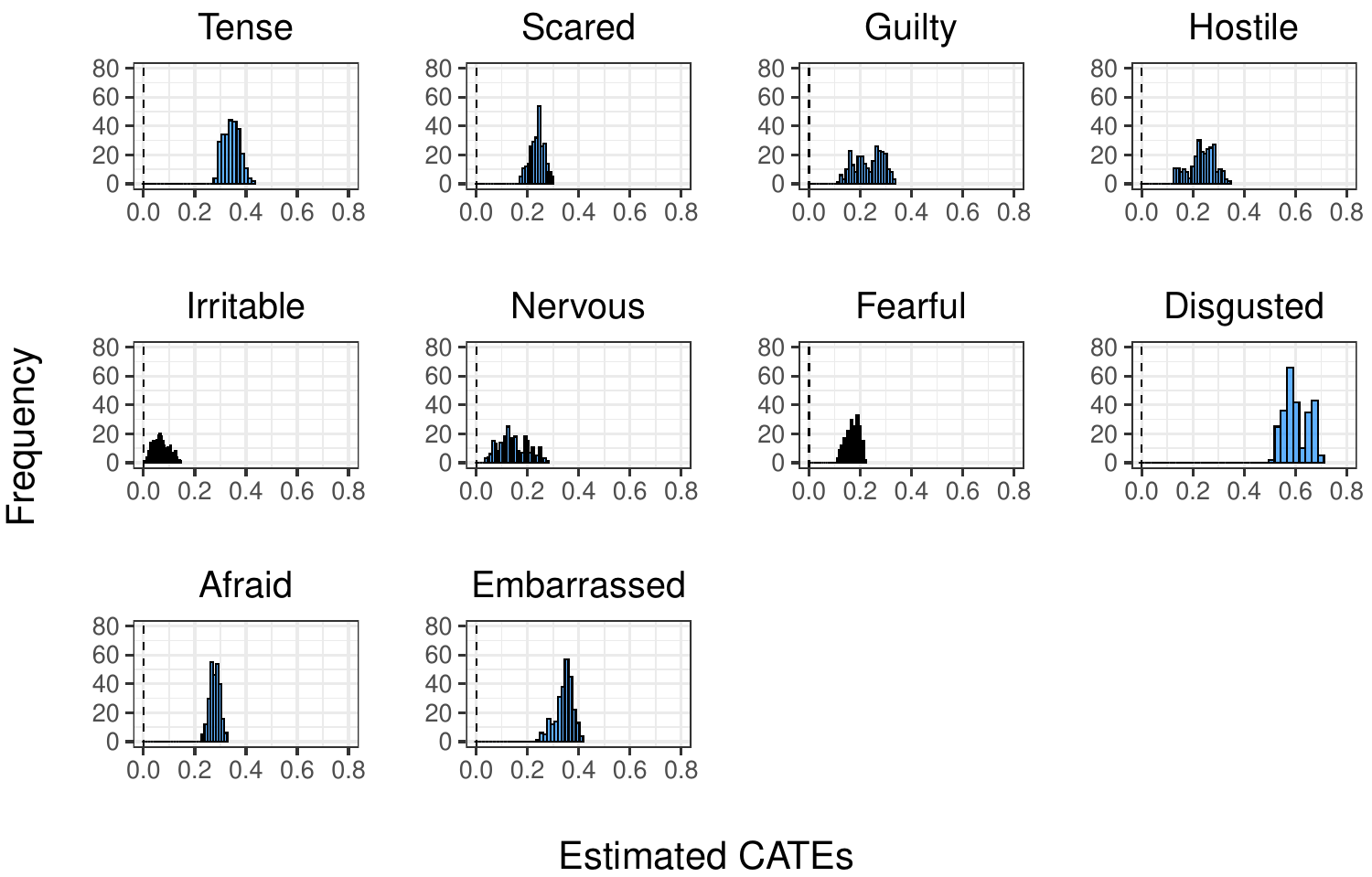}
\caption{Validating the MMR Assumption (\ref{assm:pmono}) for the effect of museum visit on democratic supports \citep{balcells2022transitional}. The figure displays a histogram for estimated CATE values for the effect of $A_i$ on $M_i$ for each of the negative emotion mediators. CATEs are estimated using causal forests \citep{wager2018estimation} with 2000 trees.}
\label{fig:balcells_cate}
\end{center}
\end{figure} 